# Where is everybody? - Wait a moment… New approach to the Fermi paradox.

I.Bezsudnov [1], A.Snarskii [2]

**The Fermi Paradox is the apparent contradiction between the high probability extraterrestrial civilizations' existence and the lack of contact with such civilizations. In general, solutions to Fermi's paradox come down to either estimation of Drake equation parameters i.e. our guesses about the potential number of extraterrestrial civilizations or simulation of civilizations development in the universe. We consider a new type of cellular automata, that allows to analyze Fermi paradox. We introduce bonus stimulation model (BS-model) of development in cellular space (Universe) of objects (Civilizations). When civilizations get in touch they stimulate development each other, increasing their life time. We discovered nonlinear threshold behaviour of total volume of civilizations in universe and on the basis of our model we built analogue of Drake equation.**

The paradox suggested by E. Fermi in 1950 has been formulated in the form of a question – «where is everybody?». At the heart of Fermi paradox is the fact of absence of observations of extraterrestrial Civilizations, their activity or traces of their activity. Let's assume that the Civilizations develop at the same rates, as unique one known to us at the Earth. In this case, it follows that *(1)* those Civilizations which have begun the development a little earlier than Earth civilization, should possess enormous technological possibilities and their activity would be conspicuous. There is a great number of publications on discussion of Fermi paradox and many different lines of thought used for its explanation *(2,3)*, in the book *(4)* fifty speculative decisions are resulted.

---

[1] Nauka - Service JSC, Moscow, Russia

[2] National Technical University of Ukraine "KPI", Dep. of General and Theoretical Physics, Kiev, Ukraine

In *(5)* Fermi paradox is discussed within the limits of the percolation theory *(6)*. The possibility of existence of infinite cluster of Civilizations (using the percolation theory terminology) is considered as the answer to Fermi paradox. It is assumed here, that Civilizations are born at the same moment and that they live infinitely long. Civilizations not belonging to such infinite cluster remain separated, lonely.

We support the view that Civilizations are being born, develop, reach their golden ages and then die. It is possible to understand as dying both destruction, loss of interest to world around them, a cutoff in the development of technologies and the terminations of interaction with surroundings. According to Lipunov *(7)*, each Civilization has a date of birth and the initial life time limited by some specific factors. Possible reason for disappearance of Civilizations is the loss of interest to development - «the universal cause of death of Intellect in the Universe can be connected with loss of its basic functions – knowledge functions» *(7)*. We assume, that the unique reason which can prolong a lifetime of the Civilization, is the contact to other Civilizations. The meeting of Civilizations generates the new purposes and objects of knowledge, necessity to use an Intellect.

We propose the following model of existence, development and dying of Civilizations in the Universe. Civilizations are born with the probability $n$ and have the initial life time (time of expansion) $T_0$ set in advance after which they die (disappear or cease to show itself). Contact of developing Civilizations increases actual life time for everyone contacted on a certain extra time, named further a time bonus $T_b$. We will refer to this model as a Bonus Stimulated model (BS-model).

Further we will use terminology adapted for discussion of Fermi paradox even though the offered model can describe and another phenomena. It is possible to offer the analogy to economy, and in this instance the initial lifetime $T_0$ is a seed capital of the new formed company. Also this capital defines time of company's independent life, while association process of this company with other companies stimulate the further development, increasing stability of existence of association by $T_b$ value etc.

**BS-model of development of Civilizations in Universe**

BS-model defines the Universe as the cellular automata. The rules of this automata define the behavior of Civilizations - objects which born, grow and die in the Universe. A Civilization is a set of the cells joined by their "history". Specific for this cellular automata is that its behavior is depended not only on its condition at the previous moment, but on the events which took place in the Universe much earlier, beginning with the births of the oldest Civilization, existing currently in the Universe. In BS-model the depth of "history" is not constant.

In the present paper the two-dimensional cellular automata of size $N \times N$ is considered. A Civilization is a square of Universe cells. The centre of square is the Civilization birthplace. Following rules are set:

1. The Civilization birth. New Civilization can be born in the free cell of Universe not occupied currently by another Civilizations. Probability of birth - $n$, size – single cell.

2. Civilization growth and dying. For each step of time each Civilization changes its size by one layer of cells on each side. If the Civilization exists less $T_0$ time steps, its size increases, and if it is more than $T_0$ - decreases. When the size becomes equal to the zero, then the Civilization is considered to disappear (this alien culture died, not demonstrating any signs of life, etc.). Initial life time $T_0$ is the same to all Civilizations.

3. Civilizations aggregation. If, during the growth, one Civilization meets another (the same cell of the Universe has to belong to both Civilizations) they form the cluster and lifetime $T_0$ of each Civilization is increased by bonus time $T_b$. If the Civilization joins the existing cluster, the bonus $T_b$ is given to all Civilizations in the cluster. Further development of each Civilization goes according to rule 2.

On Fig. 1 the development of Civilizations in Universe is shown in the space and time schematically (without «cellness»). The time axis is directed from below upwards. In this view the Civilization is two cones with joined bases and cone point specifies the birthplace of Civilization.

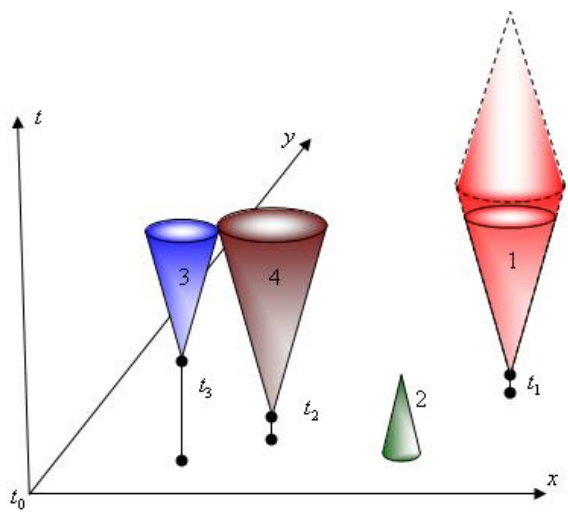

Fig. 1. Sketch of Civilizations' development in the space and time. The single Civilization 1 develops since $t_1$ to the present moment and will die at the time $2T_0$ in the future, the Civilization 2 – is born earlier, reached $t_0$ and dies. Civilizations 3 and 4 were born at $t_3$ and $t_4$ accordingly, presently they both form cluster and obtain for everyone a time bonus $T_b$, then they will develop further.

On Fig. 2 consecutive time slices of the developing Universe (fragment) are presented schematically (with «cellness»). The volume of the occupied space $V$ grows eventually. In the shown fragment: 115, 148, 197 occupied cells for each consecutive time step.

Thus, in BS-model the developing of Civilizations in Universe is governed by three parameters: $T_0$ - initial life time of the Civilization, $n$ - probability of birth of the Civilization, $T_b$ - bonus time. Incorporation of bonus time, an extra time of development (growth) of the Civilization is essentially new assumption of BS-model. Fermi paradox within the limits of the proposed model will be resolved, if it will be known the values of parameters when the Universe will be occupied by a single cluster of Civilizations or on the contrary, rare or local associations of Civilizations will take place only. In the latter case it is fair to say that Civilizations are separated and we are single in the Universe.

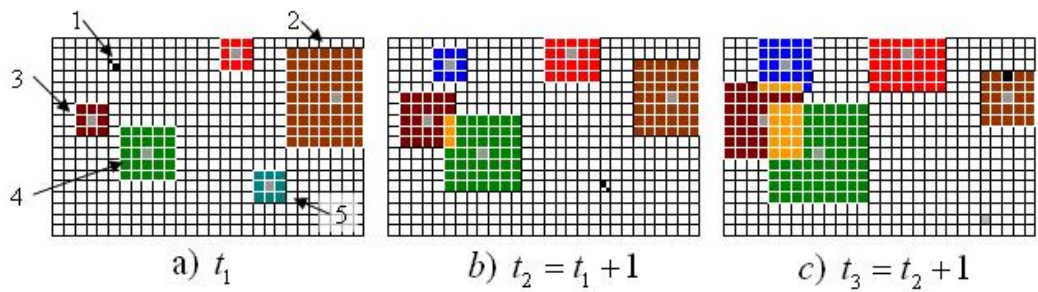

Fig. 2. Birth, growth and dying of Civilizations in Universe (three consecutive time steps). The Civilization 1 is born at $t_1$ and further develops with steps $t_2$ and $t_3$, the Civilization 2, having existed more than $T_0$ time steps, dies and goes away, Civilizations 3, 4, and then 1 on a step $t_3$ form the cluster. Lifetime for 3 and 4 will increase by $T_b$. After step $t_3$ Civilizations 3 and 4 will possess bonus $2T_b$, and 1 – bonus $T_b$. The Civilization 5 dies and disappears at step $t_3$.

The purpose of numerical simulation was to find such set of Civilization parameters $\{T_0, T_b, n\}$ at which there is a transition similar to phase transition, from the Universe with volume occupied by Civilizations $V \ll V_0$, where $V_0$ is Universe volume, to almost civilized Universe with total volume of Civilizations of the order $V_0$.

The size of cellular automata used to simulate Universe was $10^4 \times 10^4$ cells, modeling time $320\,000$ steps. Lower and upper limits for initial life time and bonus was $T_0 = 4 \div 20$ and $n = 3 \cdot 10^{-8} \div 3 \cdot 10^{-5}$ accordingly.

**Status of Universe in BS-model without bonus $T_b = 0$.**

For the case $T_b = 0$ we have to notice that the volume occupied by Civilizations $V$ will fluctuate around certain value. The cause of fluctuations is a random overlap of neighboring Civilizations that form small clusters made of independent Civilizations ($T_b = 0$).

Let us not to consider this overlapping. In this case we are able to obtain the analytical dependence of the occupied volume $V$ on Civilization parameters $\{T_0, T_b = 0, n\}$. In the Universe at each time interval there will be Civilizations of all linear sizes from 1 to $2T_0 + 1$. For all sizes except maximal one there exists a pair of Civilizations – one growing and one dying away. Therefore number of Civilizations of maximal size $2T_0 + 1$ is equal to $n \cdot N^2$, for smaller sizes $2 \cdot n \cdot N^2$ and following equation can be written

$$V/V_0 = n \cdot \left[ (2T_0 + 1)^2 + 2 \cdot \sum_{t_0=0}^{T_0-1} (2t_0 + 1)^2 \right]. \tag{1}$$

Considering in (1) that $T_0 \gg 1$ it is possible to estimate following sum as $\sum_{t_0=0}^{T_0-1} (2t_0 + 1)^2 \sim T_0^3$. Thus, from (1) one can expect that the volume occupied by Civilizations in the Universe $V$ is directly proportional to probability of birth of the Civilizations $n$ and to cube of initial life time $T_0$.

$$V \sim n^\alpha T_0^\beta, \quad \alpha = 1, \quad \beta = 3 \tag{2}$$

Numerical simulation of BS-model for values $T_0 = 4 \div 20$ and $n = 3 \cdot 10^{-8} \div 3 \cdot 10^{-5}$ gives following values for indexes $\alpha$ and $\beta$ in equation (2)

$$\alpha = 0.91 \pm 0.005, \qquad \beta = 2.79 \pm 0.02 \tag{3}$$

The $\alpha$ and $\beta$ values calculated on the base of BS-model (3) are close to those found from theoretical estimation (2) and show that overlapping of Civilizations is rather insignificant. Accordingly, the quantity of contacts appears are also insignificant and each Civilization in such Universe will live as we do now: it is impossible to assert "no, aliens is a myth", but it is impossible to prove "yes", as the contact is hardly probable.

**BS-model with bonus $T_b > 0$. The Contact of Civilizations.**

Another state of the Universe is able to reached in case of a nonzero bonus $T_b > 0$, at which Fermi paradox can be solved positively: they do exist, but it is necessary to wait.

At small bonus values $T_b$ the behavior of the Universe has a little difference from stable no-bonus state with the fluctuations, described in Sec. 3. The volume occupied by Civilizations insignificantly increases at $T_b$ increase, as well as amplitude of its fluctuations does. However at some threshold bonus $T_{bc}$ the actual life time of Civilizations at their consecutive association increases so much that allows them to develop enough to be able to form huge cluster filling almost all the Universe.

In Fig. 3 we present the dependence of relative occupied volume in the Universe $V/V_0$ upon time for $n = 1 \cdot 10^{-6}$ and $T_0 = 7$ for different $T_b$ values. Splitted Universe scenarios 1 and 2 for $T_b = 0$ и $T_b = 28$ are shown and civilized Universe scenario 3 given for $T_b = 38$.

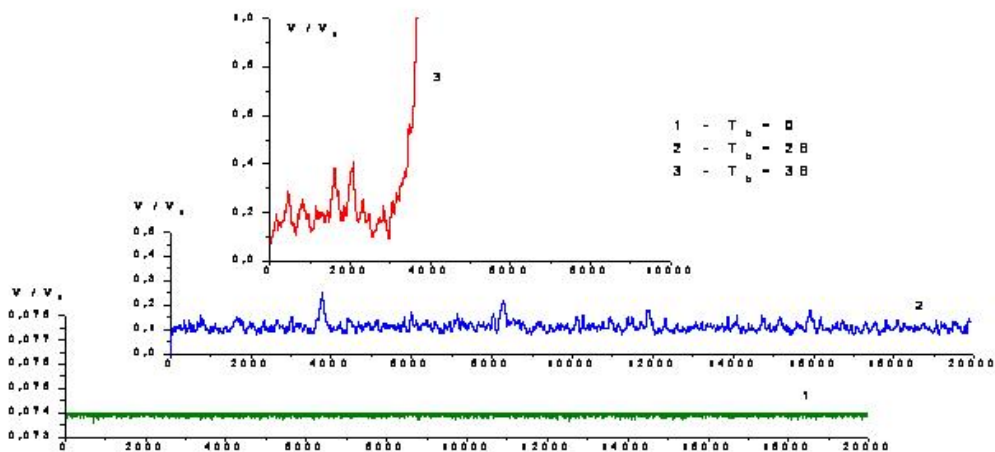

Fig. 3. Possible scenarios of the development of the Universe. The dependence of the relative occupied volume $V/V_0$ on time for various time bonuses is shown(specified in drawing).

For scenario 3 further growth of the occupied volume $V$ will continue up to $V = V_0$. Considerable difference of the occupied volume $V/V_0 \square 0,01$ at which Universe will be surely civilized in BS-model from threshold value at percolation *(5)* Universe models $V/V_0 = 0,5$ is unexpected. It shows basic difference of "phase" transition in BS-model from percolation phase transition situation.

On Fig. 4 dependence of averaged (over 100 runs) relative occupied volume $V/V_0$ for $n = 1 \cdot 10^{-6}$ and $T_0 = 7$ on a time bonus $T_b$ is presented. Results for different Universe simulation times are presented, each following consecutive time is 4 times longer than previous.

Thus, (see Fig. 4) there exists a critical bonus value $T_{bc}$ sufficient for starting the aggregation of Civilizations in Universe. One can say that critical value of a bonus $T_{bc}$ points to the beginning of "civilizationing" or "globalization" process. Simulation time does not affect $T_{bc}$ value.

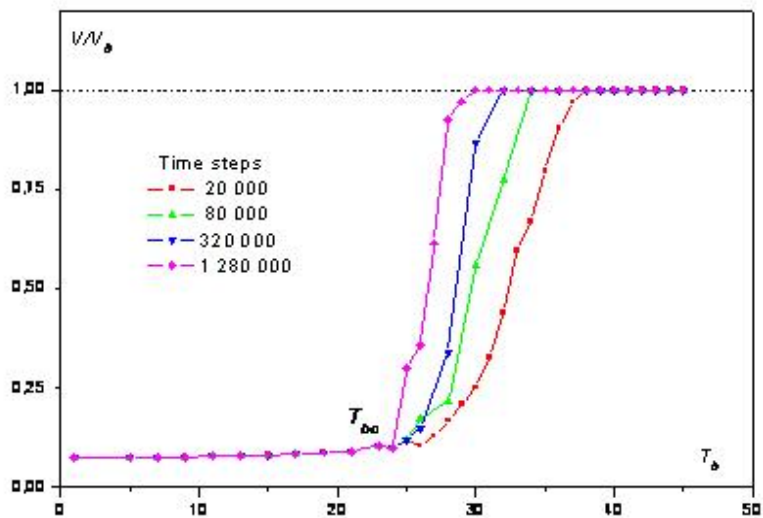

Fig. 4. Dependence of averaged relative occupied volume $V/V_0$ for $n = 1 \cdot 10^{-6}$ and $T_0 = 7$ on a time bonus $T_b$ at various simulation times (specified in drawing).

In the language of Fermi paradox we conclude that the older is Universe (longer simulation time), the more definitely it arrives at their final state – transition from split Universe to civilized one occurs certainly at sufficient $T_b \geq T_{bc}$. In the younger Universe (short simulation time) for $T_b \geq T_{bc}$ aggregation take place with some probability which, in turn, depends on Universe life time. The longer Universe exist, the greater is the probability of aggregation of Civilizations. For this "globalization" to happen there exists an obstacle, it is necessary to wait, but now our Earth Civilization has hope, hope for contact, of coarse if $T_{bc}$ will be sufficiently large for the contact.

Fig. 5 presents the final state of the Universe in BS-model, It shows an interfacial surface between the split and civilized status of Universe.

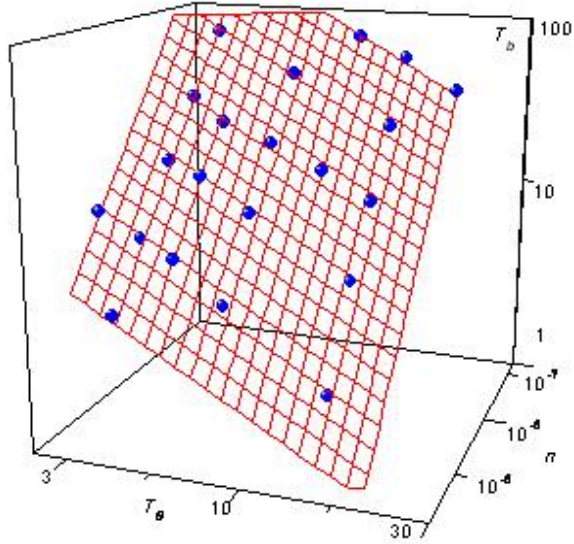

Fig. 5 Dependence of a critical time bonus $T_{bc}(T_0, n)$ on initial life time of Civilizations and probability of their birth.

Indeed, (see Fig. 5) the higher birth probability of Civilizations and the longer initial life time, the time bonus should be less to conquer the Universe by bonus stimulated Civilizations. Again we represent $T_{bc}(T_0, n)$ as power dependence on $T_0$ and $n$. We found from numerical simulations appropriate exponents (critical indexes) and coefficient:

$$T_{bc} = k \cdot n^{-\gamma} \cdot T_0^{-\lambda}, \qquad k = 0.024 \pm 0.03, \qquad \gamma = 0.62 \pm 0.01, \qquad \lambda = 0.84 \pm 0.02. \qquad (4)$$

If $T_b$ is below threshold value $T_b < T_{bc}$, we do not stand a chance to come into contact. It is the main feature of BS-model of the Universe, that the new parameter

$T_b$ changes simple structure of Drake equation *(8,9)*, that defines quantity of Civilizations with which we will be able to get into contact

Drake equation *(8,9)* depends on seven different parameters: some of them concern Universe astronomical quantities and others are Civilizations parameters $N = R \cdot f_p \cdot n_c \cdot f_l \cdot f_i \cdot f_c \cdot L$, where $N$ — number of reasonable civilizations already in contact, $R$ — the average rate of star formation per year in our galaxy, $f_p$ — the fraction of stars that have planets, $n_c$ — the average number of planets that can potentially support life per star with planets, $f_l$ — the fraction of the above that eventually produce life at some point of their development, $f_i$ — the fraction of the above that actually will develop intelligent life, $f_c$ — the fraction of civilizations that develop a technology that releases detectable signs of their existence; $L$ — the length of time that civilizations will be able to release detectable signals into space.

Modern, certainly, not indisputable, estimations for $N$ are the values of an order of 1, (more precisely *(10)* from 0.05 to 5000). The same estimation is also true for any other Civilization in our Universe, hence, Intellect should extend on all of the Universe. While this fact did not happened yet, unfortunately, or to be more precise we are not included yet in this global process which, probably, already going on!

**Analogue to Drake equation and conclusion remarks**

Drake equation *(8,9)* is widely known and is often used in discussion of Fermi paradox. It is product of various probabilities or, more likely, «improbabilities». This equation takes into account both parameters of our Universe, s well as the probabilities related to development of Civilizations in the Universe. Those last ones assume independence of Civilizations from each other, their infinitely long lifetimes and constant expansion in space. Proposed BS-model of development of Civilizations in the Universe does not contain these assumptions. Probably Drake equation has to have even more parameters and new models will be necessary to define them more precisely, but proposed BS-model shows that there can be other parameters still not considered in Drake formula.

Limited initial life time $T_0$ together with stimulating bonus time $T_b$ dramatically changes the development of Civilizations in the Universe. A consequence of three logical laws of Civilization development brings about a threshold presence, «phase transition» in status of the Universe. In the space of Civilizations parameters there exists a surface dividing a status of the Universe in two "phases": split and civilized. Otherwise, using (4) it is possible to write down the following relation:

$$T_b \cdot n^\gamma \cdot T_0^\lambda > k . \qquad (5)$$

When the set of Civilization parameters $\{T_0, T_b, n\}$ satisfies (5) the Universe will be civilized and we will have a chance to meet aliens or at least to observe them in the space. The equation (5) is analogue of Drake equation in BS-model.

Certainly, the BS-model cannot be used for an estimation of probability of Civilization contact in real Universe directly, but it offers a way to take into account non-uniformity in the development of Civilizations and their dependence on each other. As a result the scenario of development of Intellect in Universe becomes not trivial.

The BS-model is rather simple but allows for further improvements and possible rules alteration, introducing of probabilities of any events, etc. These modifications can be made both with the reference to Fermi paradox simulation, as well as to other possible applications of this model.

Fermi paradox: proposed and investigated BS-model is moderately optimistic. It is shown that there exists a scenario when at the given moment almost all Civilizations are lonely– «there is nothing», however after some, sufficiently prolong time Civilizations will get into a contact and the Universe as a whole becomes civilized. Conclusion is that it is necessary to wait!

**Acknowledgements.** Authors are indebted to Maxim Zhenirovskii and Alexander Morozovskii for helpful comments of this work and also to Vladimir Lipunov who implicitly push us to investigate intellect in Universe.